\RequirePackage{fix-cm}
\documentclass[smallextended]{svjour3}       
\smartqed  
\usepackage{amssymb}
\usepackage{amsmath}
\usepackage{graphicx}
\journalname{Ricerche di Matematica}
\begin{document}

\title{Multi-gradient fluids  \thanks{The paper is dedicated
		to Professor Tommaso Ruggeri. \\
		The work was supported by National Group of Mathematical Physics
		GNFM-INdAM (Italy).}
} 
\subtitle{}

\titlerunning{Multi-gradient fluids}        
\author{Henri Gouin}

\institute{ \at
             Aix-Marseille Univ, CNRS, Centrale Marseille, M2P2 UMR
             7340, Marseille, France. \\
             Tel.:+33-491485120\\
              \email{henri.gouin@univ-amu.fr; henri.gouin@yahoo.fr} \\ 
          ORCID iD: 0000-0003-4088-1386}
\date{Received: 25 January 2018}
\maketitle

\begin{abstract}
An internal energy function of  the mass density, the volumetric
entropy   and their gradients at $n$-order generates the
representation of \emph{multi-gradient fluids}. Thanks to
Hamilton's principle, we obtain a thermodynamical form of the
equation of motion  which generalizes the case of perfect
compressible fluids.
First integrals of flows are extended cases of perfect
 compressible fluids.
The equation of motion and the equation of energy  are written for
dissipative cases, and are compatible with the second law of
thermodynamics.
\end{abstract}

\section{Introduction}

Phase transitions between liquid and vapour are associated with
a bulk internal-energy per unit volume $\varepsilon_{0} (\rho,
\eta)$ which is non-convex function of mass density $\rho $
 and volumetric entropy $\eta$. Consequently, in continuous theories,
the simplest model allowing to study inhomogeneous fluids inside
interfacial layers considers an internal energy per unit volume
$\varepsilon$ in the form
\begin{equation}
\varepsilon  =\varepsilon_{0} (\rho,
\eta)+{\frac{1}{2}}\,\lambda\,|\mathrm{{%
grad}\rho\,|^{2},}  \label{cahnenergy}
\end{equation}
where the second term is associated with the non-uniformity of
mass density $\rho$
and $\lambda\,$ is a coefficient independent of $\eta$ and $\mathrm{{grad}%
\,\rho }$. \newline The energy form in Eq. \eqref{cahnenergy} has
been introduced by van der Waals and is widely used in the
literature \cite{Waals,Cahn}. The model describes interfaces as
diffuse layers and has many applications for describing
micro-droplets, contact-lines, nanofluidics, thin films, fluid
mixtures and
vegetal biology \cite%
{Gavrilyuk,garajeu,GouinH,GouinHR}. Thanks to the second-gradient
theory, the model was extended in continuum mechanics,
to the behaviour of  strongly  inhomogeneous  media \cite%
{Germain,Gouin-Ruggeri,Eremeyev}. Nonetheless, at equilibrium,
expression of energy given by Eq. \eqref{cahnenergy} yields an
uniform temperature in all the parts of inhomogeneous fluids
\cite{Rowlinson}. Consequently, the volumetric entropy varies with
the mass density in the same way as in the bulks; this fact leads
to monotonic variations of densities but  they are qualitative
features of non-monotonic behaviours in transition layers which
require two or more independently varying densities - entropy
included - (chapter 8 in \cite{Rowlinson}); additively, the
temperature through liquid-vapour interfaces is not necessary
constant \cite{Seppecher}. So, with the form of energy given by
Eq. \eqref{cahnenergy}, the thermodynamics of inhomogeneous fluids
is neglected and it is not possible to model flows with strong
variations of temperature such as those across non-isothermal
interfaces. In fact, it is difficult to take account of the
gradient of temperature    in the expression of the internal
energy, but we can use the gradient of entropy. The simplest model
was called thermocapillary fluids \cite{casal4}. Such a
second-gradient behaviour has also been considered by several
authors in   physical problems when  the temperature is not
constant in  inhomogeneous parts of complex media
\cite{Maitournam,Forest}.

We improve the model: when strong variations of  mass density and
entropy occur, a general case can consider  fluids when the volume
internal energy depends on
mass density, volumetric entropy and their gradients up to a convenient $n$%
-order (with $n \in \mathbb{N}$). \newline The Hamilton principle
allows to write the equation of conservative motions in an
universal thermodynamic form structurally similar to the equation
obtained in the case of conservative perfect fluids \cite{Serrin}.
Consequently,  all perfect fluid properties, as the Kelvin
theorems  are preserved and isentropic or isothermal motions can
be studied.  \newline The case of dissipative motions is
considered. From the equation of motion, the conservation of mass
and the balance of entropy, we deduce the balance of equation of
energy. The equation of energy is not in the same form as that of
classical fluids. A new vector term -- in form of   flux of energy
-- appears, which takes the gradients of mass density and entropy
into account. The Clausius-Duhem inequality can be deduced from
the viscous  dissipation and the Fourier inequality: that proves
multi-gradient fluids are compatible with  the second law of
thermodynamics.

\emph{Notations:} For any vectors $\boldsymbol{a,b}$, term $\boldsymbol{a}%
^{T }\boldsymbol{b}$ denotes the scalar product (line vector
$\boldsymbol{a}^{T}$
is multiplied by column vector $\boldsymbol{b}$) and tensor $\boldsymbol{a} {%
\ }\boldsymbol{b}^{T }$ (or $\boldsymbol{a}\otimes \boldsymbol{b}$) denotes
the product of column vector $\boldsymbol{a}$ by line vector $\boldsymbol{b}%
^{T},$ where superscript $^T$ denotes the transposition. Tensor $\boldsymbol{%
I}$ denotes the identity transformation.

 \emph{Principal nomenclatures:}
 \newline
 $\rho$, $\eta$ and $s$  denote the mass
 density,   volumetric entropy and specific entropy. \newline
 $\varepsilon$ denotes the volume internal energy. \newline
 $\tilde\mu$ and $\tilde T$ denote the  extended chemical
 potential   and  extended temperature.\newline
 ${\mathcal{D}}_{t}$ denotes the space  occupied by
the fluid at time $t$.\newline
 ${\mathcal{D}}_{0}$ denotes the reference space.\newline
$\boldsymbol{F}$ denotes the deformation gradient.\newline
$\boldsymbol{u}$ and $\boldsymbol{\gamma}$   denote the velocity
and acceleration of the fluid.\newline  $\Xi$, $\theta $ and $H$
denote the generalized  chemical potential,    generalized
temperature  and generalized enthalpy.
\section{Model of multi-gradient fluids}

 Perfect fluids of grade $n+1$ ($n \in \mathbb{N}$)
are continuous media with a volume internal energy $\varepsilon $
which is a function of $\rho $ and $\eta $ and their gradients in
form
\begin{equation}
\varepsilon =\varepsilon (\rho ,\nabla \rho ,{\ldots },\nabla
^{p}\rho, {\ldots },\nabla ^{n}\rho, \, \eta , \nabla \eta
,{\ldots },\nabla ^{p}\eta, {\ldots },\nabla ^{n}\eta  )
,\label{Intenergy}
\end{equation}%
where $\nabla ^{p}$, $p\in \{1,\ldots ,n\}$, denotes the
successive gradients in   space ${\mathcal{D}}_{t}$ occupied by
the fluid at time $t$.
\begin{equation*}
 \mathrm{grad}^p\,
\rho\equiv\nabla ^{p}\rho =\left\{\rho\,,_{ x_{j_1}} \ldots
,_{x_{j_p}}\right\} \quad \mathrm{and}\quad \mathrm{grad}^p\,
\eta\equiv\nabla ^{p}\eta =\left\{\eta\,,_{ x_{j_1}} \ldots
,_{x_{j_p}}\right\},
\end{equation*}%
where the subscript ``comma'' indicates the partial derivative and
$x_{j_1} \ldots x_{j_p} $ belong to components of Euler variables
$\boldsymbol{x}= \left [ x_{1},x_{2},x_{3}\right]^T  \in
{\mathcal{D}}_{t}$, space of the fluid at time $t$. We indicate
indices in subscript position without taking account of the tensor
covariance or contravariance. We deduce
\begin{eqnarray}
d\varepsilon =\frac{\partial \varepsilon }{\partial \rho }\,d\rho +\frac{%
\partial \varepsilon }{\partial \eta }\,d\eta +\left(\frac{\partial \varepsilon }{%
\partial \nabla \rho }\, \vdots\, d\nabla \rho\right) +\left(\frac{\partial \varepsilon }{%
\partial \nabla \eta }\, \vdots\, d\nabla \eta \right) \notag \\
+ \ldots + \left(\frac{\partial \varepsilon
}{\partial\nabla^{n}\rho }\, \vdots\, d\nabla ^{n}\rho\right)
+\left(\frac{\partial \varepsilon }{\partial \nabla ^{n}\eta }\,
\vdots\, d\nabla ^{n}\eta\right) . \label{differentielenergy}
\end{eqnarray}
Sign $\  \vdots\  $ means the complete contraction of tensors such
that,
\begin{equation*}
\left(\frac{\partial \varepsilon }{\partial \nabla ^{p}\rho }\, \vdots\,
d\nabla
^{p}\rho\right) \equiv\left( d\nabla ^{p}\rho \, \vdots\, \frac{\partial \varepsilon }{%
\partial \nabla ^{p}\rho} \right) =  \varepsilon ,_{\rho,_{x_{j_1}} \ldots
,_{x_{j_p}}} d\rho,_{x_{j_1}} \ldots ,_{x_{j_p}},
\end{equation*}
\begin{equation*}
\left(\frac{\partial \varepsilon }{\partial \nabla ^{p}\eta }\, \vdots\,
d\nabla
^{p}\eta\right) \equiv\left(d\nabla ^{p}\eta \, \vdots\, \frac{\partial \varepsilon }{%
\partial \nabla ^{p}\eta }\right) =   \varepsilon ,_{\eta,_{x_{j_1}} \ldots
,_{x_{j_p}}} d\eta,_{x_{j_1}} \ldots ,_{x_{j_p}} ,
\end{equation*}
where repeated subscripts correspond to the summation (\footnote{%
Due to the fact that $\varepsilon ,_{\rho,_{x_{j_1}} \ldots
,_{x_{j_p}}} d\rho,_{x_{j_1}} \ldots ,_{x_{j_p}}= d\rho,_{x_{j_1}}
\ldots ,_{x_{j_p}}\varepsilon,_{\,\rho,_{x_{j_1}} \ldots
,_{x_{j_p}}}$ and $\varepsilon ,_{\eta,_{x_{j_1}} \ldots
,_{x_{j_p}}} d\eta,_{x_{j_1}} \ldots ,_{x_{j_p}}= d\eta,_{x_{j_1}}
\ldots ,_{x_{j_p}}\varepsilon,_{\,\eta,_{x_{j_1}} \ldots
,_{x_{j_p}}}$, we indifferently permute the position of the two
terms in the summation.}). We call the thermodynamical functions,
\begin{equation*}
\qquad \tilde\mu=\frac{\partial \varepsilon }{\partial \rho }\quad
\mathrm{and}\quad \tilde T=\frac{\partial \varepsilon }{\partial
\eta }\;,
\end{equation*}
 the \emph{extended chemical potential} and the \emph{extended temperature},
respectively. \newline We may also consider the case of an
inhomogeneous configuration in the reference space by taking the
reference position of the fluid into account; for the sake of
simplicity, we will not do it. So, the fluid is supposed to have
infinitely short memory and the motion history until an
arbitrarily chosen past does not affect the determination of
stresses at present time.

\section{Equation of motions. Generalized temperature and  generalized chemical potential}

\subsection{Virtual motions}

The virtual power principle is a convenient way to obtain the
motion equation. For conservative motions, it writes as the
Hamilton principle. A particle is identified in Lagrange
representation by position $\boldsymbol{X}=\left[X_{1},X_{2},X_{3}\right]^T  \in  {{%
\mathcal{D}}}_{0} $, reference space
 of the fluid. The variations of a motion are
deduced from the functional family of virtual motions
\begin{equation*}
\boldsymbol{X}=\mathbf{\psi }(\boldsymbol{x},t;\beta ),
\end{equation*}%
where $\boldsymbol{x}\in \mathcal{D}_{t}$,  $\beta $ is a scalar
parameter defined in the vicinity of $\, 0$, when the real motion
corresponds to $\beta =0$. The virtual displacement $\delta$
associated with any variation of the real motion is written in the
form
\begin{equation}
\delta \boldsymbol{X}=\left. \frac{\partial \mathbf{\psi (\boldsymbol{x}}%
,t;\beta )}{\partial \beta }\right\vert _{\beta =0}.
\label{virtual displacements}
\end{equation}%
The variation is dual and mathematically equivalent to Serrin's one (\cite%
{Serrin}, p. 145). Additively, we only consider variations with compact
support in $D_{0}$; consequently, $\delta \boldsymbol{X}=0$ on boundary $%
\partial {\mathcal{D}}_{0}$ of ${\mathcal{D}}_{0}$.\newline
The mass density satisfies the mass conservation law,
\begin{equation}
\rho \,\det \boldsymbol{F} =\rho _{0}(\boldsymbol{X}) ,  \label{density}
\end{equation}%
where $\rho _{0}$ is defined on ${{\mathcal{D}}}_{0}$ and
$\boldsymbol{F} =\partial\boldsymbol{x}/\partial\boldsymbol{X}$ is
the deformation gradient. Equivalently, we can write
\begin{equation*}
\frac{\partial \rho }{\partial t}+\mathrm{{div}\left( \rho\, \boldsymbol{u}%
\right) =0 ,}
\end{equation*}%
where $\boldsymbol{u}$ is the fluid velocity. The motion is
supposed to be conservative and the specific entropy $s=\eta /\rho
$ is constant along each trajectory (isentropic motion)
\begin{equation}
s =s_{0}(\boldsymbol{X}),  \label{specificentropy}
\end{equation}%
or
\begin{equation*}
\eta \,\det \boldsymbol{F}=\eta _{0}(\boldsymbol{X})\qquad \mathrm{with}%
\qquad \eta _{0}(\boldsymbol{X})=\rho _{0}(\boldsymbol{X})\ s_{0}(%
\boldsymbol{X}).
\end{equation*}
Equivalently to relation \eqref{specificentropy}, we can write
\begin{equation*}
\frac{\partial \eta }{\partial t}+\mathrm{{div}\left( \eta\, \boldsymbol{u}%
\right) =0 .}
\end{equation*}

\emph{\textbf{Lemma}}: \emph{The variations of mass density and
volumetric entropy verify}
\begin{equation}
\delta \rho =\frac{\mathrm{{div}_{0}\left( \rho _{0}\ \delta \boldsymbol{X}%
\right) }}{\det \boldsymbol{F}}\qquad \mathrm{and}\qquad \delta \eta =\frac{%
\mathrm{{div}_{0}\left( \eta _{0}\ \delta \boldsymbol{X}\right) }}{\det
\boldsymbol{F}} \, ,  \label{variation density entropy}
\end{equation}%
\emph{where} ${\rm div}_{0}$ \emph{denotes the divergence in
reference space} $D_{0}$. \newline

\noindent The proof of the lemma comes from the velocity
definition,
\begin{equation*}
\frac{\partial \boldsymbol{X}(\boldsymbol{x},t)}{\partial
\boldsymbol{x}}\, \boldsymbol{u}+\frac{\partial
\boldsymbol{X}(\boldsymbol{x},t)}{\partial t} = 0 ,
\end{equation*}
which implies
\begin{equation}
\frac{\partial \boldsymbol{X}}{\partial \boldsymbol{x}}\,\delta
\boldsymbol{u}+\frac{\partial \delta \boldsymbol{X}}{\partial \boldsymbol{x}}\,\boldsymbol{u%
}+ \frac{\partial \delta \boldsymbol{X}}{\partial t%
}=0\qquad {\rm or} \qquad \delta \boldsymbol{u}= -\boldsymbol{F}%
\, \frac{d\left( {\delta \boldsymbol{X}}\right) \label{velocity} }{dt},
\end{equation}
where $d/dt$ denotes the material derivative. But,
\begin{equation*}
\boldsymbol{F\ F}^{-1}=\  \boldsymbol{%
I} \quad \Longrightarrow\quad \delta
\boldsymbol{F=-F\ \delta F}^{-1}\boldsymbol{F=-}\frac{\partial \boldsymbol{x}%
}{\partial \boldsymbol{X}}\frac{\partial \delta \boldsymbol{X}}{\partial
\boldsymbol{X}}.
\end{equation*}%
From the Jacobi identity and Eq. \eqref{density}, we get
\begin{equation*}
\delta \det \boldsymbol{F=}-\det \boldsymbol{F}\ \text{tr}\left( \frac{%
\partial \delta \boldsymbol{X}}{\partial \boldsymbol{X}}\right) =-\det
\boldsymbol{F}\ \mathrm{div}_{0}\,\delta \boldsymbol{X} \quad \mathrm{and}%
\quad \rho \, =\frac{\rho _{0}(\boldsymbol{X})}{\det \boldsymbol{F}} ,
\end{equation*}
which implies
\begin{equation*}
\delta \rho =\frac{1}{\det \boldsymbol{F}}\left[\rho _{0}\, \mathrm{div}
_{0}\left( \delta \boldsymbol{X}\right) +\frac{\partial \rho _{0}}{\partial
\boldsymbol{X}}\,\delta \boldsymbol{X}\right]
\end{equation*}
and consequently relation \eqref{variation density entropy}$^1$. The same
calculus is suitable to volumetric entropy $\eta $ and we obtain relation %
\eqref{variation density entropy}$^2$.\newline \noindent Due to
definition \eqref{virtual displacements}, variation $\delta $ and
space gradients are independent, so they commute  ($\delta \nabla
^{p}\rho =\nabla ^{p}\delta
\rho $, $\delta \nabla ^{p}\eta =\nabla ^{p}\delta \eta $).  Expression %
\eqref{differentielenergy} implies,%
\begin{eqnarray}
\delta \varepsilon &=&\frac{\partial \varepsilon }{\partial \rho
}\,\delta
\rho \,+\frac{\partial \varepsilon }{\partial \eta}\,\delta \eta +\left(\frac{%
\partial \varepsilon }{\partial \nabla \rho }\,  \vdots   \,\nabla \delta
\rho\right) +\left(\frac{\partial \varepsilon }{\partial \nabla
\eta }\,  \vdots
\,\nabla \delta \eta \right)+{\ldots } \notag\\
&& + \left(\frac{\partial \varepsilon }{\partial \nabla ^{n}\rho
}\,  \vdots   \,\nabla ^{n}\delta \rho\right)
+\left(\frac{\partial \varepsilon }{\partial \nabla ^{n}\eta }\,
\vdots  \,\nabla ^{n}\delta \eta\right) \label{deltaespsilon}.
\end{eqnarray}%
We define operator $\ \mathrm{div}_{p}\ $ as follows :
\begin{equation*}
\mathrm{div}_{p}(b_{{j_{1}\ldots j_{p}}})=\left( b_{{j_{1}\ldots j_{p}}%
}\right)_{{,x_{j_{1}}\ldots ,x_{j_{1}}}},\ p\in \mathbb{N}\quad \mathrm{with%
}\quad {x_{j_{1}},\ldots ,x_{j_{p}}}\in \left\{ x_{1},x_{2},x_{3}\right\} .
\end{equation*}%
Classically, term $\left( b_{{j_{1}\ldots j_{p}}}\right) _{{,x_{j_{1}}\ldots
,x_{j_{p}}}}$ corresponds to the summation on the repeated indices $%
j_{1}\ldots j_{p}$ of the consecutive derivatives of $b_{j_{1}}\ldots j_{p}$
with respect to $x_{j_{1}}\ldots x_{j_{p}}$ (\footnote{%
For example, when \ \ $A=\left[
\begin{array}{ccc}
a_{11}, & a_{12}, & a_{13} \\
a_{21}, & a_{22}, & a_{23} \\
a_{31}, & a_{32}, & a_{33}%
\end{array}%
\right] ,$ then\newline
$\mathrm{div}A=\big[%
a_{11},_{x_{1}}+a_{21},_{x_{2}}+a_{31},_{x_{3}},\,
a_{12},_{x_{1}}+a_{22},_{x_{2}}+
a_{32},_{x_{3}},\, a_{13},_{x_{1}}+a_{23},_{x_{2}}+a_{33},_{x_{3}}%
\big],\ \mathrm{and}$ \newline
$\mathrm{div}_{2}\,A={a_{11}}_{,x_{1},x_{1}}+{a_{21}}_{,x_{2},x_{1}}+{a_{31}}%
_{,x_{3},x_{1}}+{a_{12}}_{,x_{1},x_{2}}+{a_{22}}_{,x_{2},x_{2}}+{a_{32}}%
_{,x_{3},x_{2}}+{a_{13}}_{,x_{1},x_{3}}+{a_{23}}_{,x_{2},x_{3}}+{a_{33}}%
_{,x_{3},x_{3}}\,.$}).\newline
Operator $\mathrm{div}_{p}$ is the \emph{extended divergence} \emph{at order }$p$%
. Operator ${\mathrm{div}}_{p}$ decreases from   $p$\, the tensor order and term $%
\nabla ^{p}$ increases from  $p$\, the tensor order. We deduce,%
\begin{eqnarray*}
\left(\frac{\partial \varepsilon }{\partial \nabla \rho }\, \vdots \,
\nabla \delta \rho\right) &=&\mathrm{div}\left( \frac{\partial
\varepsilon }{\partial \nabla \rho
} \, \delta \rho \right) -\mathrm{div}\left( \frac{\partial \varepsilon }{%
\partial \nabla \rho }\right)  \delta \rho \\
\left(\frac{\partial \varepsilon }{\partial \nabla ^{2}\rho }\, \vdots
\, \nabla ^{2}\delta \rho\right) &=&\mathrm{div}\left[
\left(\frac{\partial \varepsilon }{\partial
\nabla ^{2}\rho }\, \vdots \, \nabla \delta \rho\right) -\mathrm{div}\left( \frac{%
\partial \varepsilon }{\partial \nabla ^{2}\rho }\right) \delta \rho \right]
+\mathrm{div}_{2}\left( \frac{\partial \varepsilon }{\partial \nabla
^{2}\rho }\right)  \delta \rho \\
&\, \vdots \, & \\
\left(\frac{\partial \varepsilon }{\partial \nabla ^{n}\rho }\, \vdots
\, \nabla
^{n}\delta \rho\right) &=&\mathrm{div}  A_{n} +(-1)^{n}\mathrm{div}%
_{n}\left( \frac{\partial \varepsilon }{\partial \nabla ^{n}\rho
}\right)  \delta \rho
\end{eqnarray*}%
with  $A_{p}, \ p\in \{1,\ldots,n\}$ verifies
\begin{eqnarray*}
A_{p} &=&\left(\frac{\partial \varepsilon }{\partial \nabla
^{p}\rho }\, \vdots \,
\nabla ^{p-1}\delta \rho\right) -\left(\mathrm{{div}}\left( \frac{\partial \varepsilon }{%
\partial \nabla ^{p}\rho }\right) \, \vdots \, \nabla ^{p-2}\delta \rho\right) + %
\ldots \label{Ap}  \\
+&{(-1)}^{q-1}&\left(\mathrm{div}_{q-1}\left( \frac{\partial
\varepsilon }{\partial
\nabla ^{p}\rho }\right) \, \vdots \, \nabla ^{p-q}\delta \rho\right) +{\ldots +(-1)}%
^{p-1}\mathrm{div}_{p-1} \left(\frac{\partial \varepsilon
}{\partial \nabla ^{p}\rho }\right)  \delta \rho.\notag
\end{eqnarray*}
We obtain the same relation for the volumetric entropy,
\begin{eqnarray*}
\left(\frac{\partial \varepsilon }{\partial \nabla \eta }\, \vdots \,
\nabla \delta \eta\right) &=&\mathrm{div}\left( \frac{\partial
\varepsilon }{\partial \nabla \eta
}\,\delta \eta \right) -\mathrm{div}\left( \frac{\partial \varepsilon }{%
\partial \nabla \eta }\right)   \delta \eta \\
\left(\frac{\partial \varepsilon }{\partial \nabla ^{2}\eta }\, \vdots
\, \nabla ^{2}\delta \eta\right) &=&\mathrm{div}\left[
\left(\frac{\partial \varepsilon }{\partial
\nabla ^{2}\eta }\, \vdots \, \nabla \delta \eta\right) -\mathrm{div}\left( \frac{%
\partial \varepsilon }{\partial \nabla ^{2}\eta }\right)  \delta \eta \right]
+\mathrm{div}_{2}\left( \frac{\partial \varepsilon }{\partial
\nabla
^{2}\eta }\right) \delta \eta \\
&\, \vdots \, & \\
\left(\frac{\partial \varepsilon }{\partial \nabla ^{n}\eta }\, \vdots
\, \nabla
^{n}\delta \eta\right) &=&\mathrm{div} B_{n}  +(-1)^{n}\mathrm{div}%
_{n}\left( \frac{\partial \varepsilon }{\partial \nabla ^{n}\eta
}\right) \delta \eta
\end{eqnarray*}%
with  $B_{p}, \ p\in \{1,\ldots,n\}$ verifies
\begin{eqnarray*}
B_{p} &=&\left(\frac{\partial \varepsilon }{\partial \nabla
^{p}\eta }\, \vdots \,
\nabla ^{p-1}\delta \eta\right) -\left(\mathrm{{div}}\left( \frac{\partial \varepsilon }{%
\partial \nabla ^{p}\eta }\right) \, \vdots \, \nabla ^{p-2}\delta
\eta\right)
+
\ldots  \label{Bp}\\
+&{(-1)}^{q-1}&\left(\mathrm{div}_{q-1}\left( \frac{\partial
\varepsilon }{\partial
\nabla ^{p}\eta }\right) \, \vdots \, \nabla ^{p-q}\delta \eta \right)+{\ldots +(-1)}%
^{p-1}\mathrm{div}_{p-1}\left( \frac{\partial \varepsilon
}{\partial \nabla ^{p}\eta }\right) \delta \eta.\notag
\end{eqnarray*}
We denote
\begin{equation}
\left\{
\begin{array}{c}
 \quad\displaystyle\Xi =\tilde \mu
-\mathrm{{div}\,\boldsymbol{\Phi
}}_{1}+\mathrm{div}%
_{2}\,{\boldsymbol {\Phi }}_{2}+{\ldots +}(-1)^{n}\mathrm{div}_{n}{\,\boldsymbol{%
\Phi }}_{n},\\
\\
\quad\displaystyle\theta =\tilde T-\mathrm{{div}\,\boldsymbol{\Psi
}}_{1}+\mathrm{div}
_{2}\,{\boldsymbol{\Psi }}_{2}+{\ldots +}(-1)^{n}\mathrm{div}_{n}\,{\boldsymbol{%
\Psi }}_{n},
\end{array}%
\right.  \label{tempchem}
\end{equation}%
\newline
with
\begin{equation}
\boldsymbol{\Phi }_{p}=\frac{\partial \varepsilon }{\partial
\nabla ^{p}\rho }=\left\{ \varepsilon ,_{\rho ,_{x_{j_{1}}}\ldots
,_{x_{j_{p}}}}\right\},\qquad \boldsymbol{\Psi
}_{p}=\frac{\partial \varepsilon }{\partial \nabla ^{p}\eta
}=\left\{ \varepsilon ,_{\eta \,,_{x_{j_{1}}}\ldots
,_{x_{j_{p}}}}\right\} , \label{PhipPsip}
\end{equation}%
where ${x_{j_{1}},\ldots ,x_{j_{p}}}\in \left\{ x_{1},x_{2},x_{3}\right\} $.
We call   $\Xi\, $ and  $\, \theta $ the \emph{%
generalized chemical potential} and  the \emph{generalized
temperature},  respectively.

\subsection{The Hamilton action. Generalized enthalpy}

Let $\mathcal{L}$ be the Lagrangian of the fluid, i.e.
\begin{equation*}
\mathcal{L}=\frac{1}{2}\,\rho \,\boldsymbol{u}^{T}\boldsymbol{u}-\varepsilon
-\rho \,\Omega ,\quad \mathrm{with}\quad \boldsymbol{u}^{T}\boldsymbol{u}=|%
\boldsymbol{u}|\,^{2},
\end{equation*}%
where $\varepsilon$ is defined by Eq. \eqref{Intenergy},
$\boldsymbol{u}$ denotes the particle velocity and $\Omega$,
function of $(\boldsymbol{x},t)$, denotes the potential of external
forces. Between times $t_{1}$ and $t_{2}$, the Hamilton action
writes
\begin{equation*}
a=\int_{t_{1}}^{t_{2}}\int_{{{\mathcal{D}}}_{t}}\mathcal{L}~dx\,dt,
\end{equation*}%
where $dx\,dt$ denotes the volume element in time-space
$[t_{1},t_{2}]\times {D}_{t}$. Classical methods of variation
calculus yield the variation of Hamilton's action
\begin{equation*}
\delta a=a^{\prime }(\beta )_{|{\beta =0}}.
\end{equation*}%
From $\delta \Omega =0$, we deduce
\begin{equation*}
\delta a=\int_{t_{1}}^{t_{2}}\int_{{{\mathcal{D}}}_{t}}\left[
\rho\,
\boldsymbol{u}^{T}\delta \boldsymbol{u}+\left( \frac{1}{2}\,\boldsymbol{u}%
^{T}\boldsymbol{u}-\Omega \right) \delta \rho -\delta \varepsilon
\right] dx\,dt.
\end{equation*}%
From eqs.  \eqref{deltaespsilon} to \eqref{PhipPsip}, we deduce
the value of $\delta\varepsilon$ and
\begin{eqnarray*}
\delta a &=&\int_{t_{1}}^{t_{2}}\int_{{{\mathcal{D}}}_{t}}
\left\{\left[ \rho\,
\boldsymbol{u}^{T}\delta \boldsymbol{u}+\left( \frac{1}{2}\,\boldsymbol{u}%
^{T}\boldsymbol{u}-\Xi -\Omega \right) \delta \rho -\theta \,\delta \eta %
\right]\right. \\
&& +\ \mathrm{{div}  \left( A_{1}+{\ldots +}A_{n}+B_{1}+{\ldots
+}B_{n}\right) \Big\}}\, dx\,dt,
\end{eqnarray*}%
where term $\mathrm{{div}\left( A_{1}+{\ldots +}A_{n}+B_{1}+{\ldots +}B_{n}%
\right) }$ can be integrated on $\partial {D}_{t}$ and
consequently has a zero contribution. Then,
\begin{equation*}
\delta a = \int_{t_{1}}^{t_{2}}\int_{{{\mathcal{D}}}_{0}}\left[ \rho _{0}\,\boldsymbol{u%
}^{T}\delta \boldsymbol{u}+\det \boldsymbol{F}\left( \frac{1}{2}\,%
\boldsymbol{u}^{T}\boldsymbol{u}-\Xi -\Omega \right) \delta \rho
-(\det \boldsymbol{F})\,\theta \,\delta \eta \right] dX dt,
\end{equation*}%
where $dX$ denotes the volume element in ${D}_{0}$. Let us denote $m=\displaystyle\frac{1}{2}\,\boldsymbol{u}^{T}\boldsymbol{u}%
-\Xi -\Omega $; taking account of Eqs. \eqref{variation density
entropy} and \eqref{velocity}, we obtain
\begin{eqnarray*}
\delta a &=&\int_{t_{1}}^{t_{2}}\int_{{{\mathcal{D}}}_{0}}\left[
-\rho _{0}\,
\boldsymbol{u}^{T}\boldsymbol{F}\,\frac{d\delta \boldsymbol{X}}{dt}%
  +  (\det \boldsymbol{F})\  m\ \delta \rho -(\det \boldsymbol{F})\ %
\theta \,\delta \eta \right] dX dt= \\
&&\int_{t_{1}}^{t_{2}}\int_{{{\mathcal{D}}}_{0}}\left\{\left[ \rho _{0}\frac{%
d\left( \boldsymbol{u}^{T}\boldsymbol{F}\right) }{dt}\ \boldsymbol{-}\rho
_{0}\, \nabla _{0}^{T}m\ +\eta _{0} \,\nabla _{0}^{T}\theta %
\right] \delta \boldsymbol{X}\right\}dX dt+ \\
&&\int_{t_{1}}^{t_{2}}\int_{{{\mathcal{D}}}_{0}}\left[ -\frac{d\left( \rho
_{0}\,\boldsymbol{u}^{T}\boldsymbol{F}\delta \boldsymbol{X}\right) }{dt}+%
\mathrm{div}_{0}\left( m\,\rho _{0}\,\delta \boldsymbol{X}\right) -\mathrm{%
div}_{0}\left( \theta \,\eta _{0}\,\delta \boldsymbol{X}\right)
\right] dX dt,
\end{eqnarray*}%
where $\nabla _{0}$ and $\mathrm{div}_{0}$ denotes the gradient and
divergence operators in ${\mathcal{D}}_{0}$, respectively. The last integral
can be integrated on the boundary of $\left[ t_{1},t_{2}\right] \times {D}%
_{0}$ and consequently is null. Then,
\begin{equation*}
\delta a=\int_{t_{1}}^{t_{2}}\int_{{{\mathcal{D}}}_{0}}\left\{\left[ \rho _{0}\frac{%
d\left( \boldsymbol{u}^{T}\boldsymbol{F}\right) }{dt}\
\boldsymbol{-}\rho _{0}\, \nabla _{0}^{T}m\ +\eta _{0}\,\nabla
_{0}^{T}\theta \right] \delta \boldsymbol{X}\right\}dX dt.
\end{equation*}%
The fundamental lemma of variation calculus and $\eta _{0}=\rho
_{0}\,s_{0}$ yield
\begin{equation*}
\frac{d\left( \boldsymbol{u}^{T}\boldsymbol{F}\right) }{dt}\,\boldsymbol{-}%
\,\nabla _{0}^{T}m+s_{0}\,\nabla _{0}^{T}\theta =0.
\end{equation*}%
Due to%
\begin{equation*}
\frac{d\left( \boldsymbol{u}^{T}\boldsymbol{F}\right) }{dt}=\left(
\boldsymbol{\gamma }^{T}+\boldsymbol{u}^{T}\frac{\partial \boldsymbol{u}}{%
\partial \boldsymbol{x}}\right) \boldsymbol{F},
\end{equation*}%
where $\boldsymbol{\gamma }$ denotes the acceleration vector,  we
obtain the equation of conservative motions,
\begin{equation}
{\boldsymbol{\gamma }}+{\mathrm{{grad}}\left( \Xi +\Omega \right)
+s\,{\rm grad}\,\theta =0\quad or\quad {{\boldsymbol{\gamma
}}}+{\rm grad}\left(H+\Omega \right) -\theta \ {\rm grad}\,s=0,}
\label{motion}
\end{equation}%
where $H$ = $\Xi +s\,\theta $ denotes the\emph{\ generalized
enthalpy}. Relation \eqref{motion} is a thermodynamic form of the
equation of isentropic
motions for perfect fluids which generalizes relation (29.8) in \cite%
{Serrin}.

\subsection{Isothermal motions}

We assume that
\begin{equation}
\int_{t_{1}}^{t_{2}}\int_{{{\mathcal{D}}}_{t}}\eta ~ dx\,dt=S_{0},
\label{totalentropy}
\end{equation}
where $S_0$ is a constant. Only mass conservation \eqref{density}
is considered. The volumetric entropy
becomes an independent variable which is subject to  constraint %
\eqref{totalentropy} and action $a$ can be replaced by
\begin{equation*}
b=\int_{t_{1}}^{t_{2}}\int_{{{\mathcal{D}}}_{t}}\left( \mathcal{L}
+\theta _{0}\,\eta ~\right) dx\,dt,
\end{equation*}%
where scalar $\theta _{0}$ is a constant Lagrange multiplier
associated with integral constraint \eqref{totalentropy}.\newline
For any variation $\kappa $ of $\eta $,  with $\delta \boldsymbol{X}%
=0 $, we immediately deduce
\begin{equation*}
\ \int_{t_{1}}^{t_{2}}\int_{{{\mathcal{D}}}_{0}} \big[\det \boldsymbol{%
F}\  (\theta_{0} -\theta) \, \kappa\,\big] dX dt=0
\end{equation*}%
and consequently
\begin{equation*}
\theta =\theta _{0} ,
\end{equation*}%
that corresponds to isothermal motions in the sense of
\emph{generalized temperature} $\theta$.  When $%
\kappa =0$, for any variation of $\delta \boldsymbol{X}$,  we get
the equation of motion in the form
\begin{equation*}
\boldsymbol{\gamma}+\mathrm{{grad} \left( \Xi +\Omega \right) = 0
.}
\end{equation*}
At equilibrium, without body forces,
\begin{equation*}
\Xi =\mu _{0}\quad \mathrm{and}\quad \theta =\theta _{0}\, ,
\end{equation*}%
where $\mu _{0}$ denotes the chemical-potential value in fluid bulks.

\subsection{Conservative properties of perfect multi-gradient fluids}

The circulation of velocity vector $\boldsymbol{u}$ on a closed fluid-curve $%
\mathcal{C}$ is $\displaystyle\mathcal{J }=\oint_{\mathcal{C}} \boldsymbol{u}%
^T\, d\boldsymbol{x}$. From \cite{Serrin} p. 162,
\begin{equation*}
\frac{d}{dt}\oint_{\mathcal{C}} \boldsymbol{u}^T\, d\boldsymbol{x}%
=\oint_{\mathcal{C}} \boldsymbol{\gamma}^T\, d\boldsymbol{x}
\end{equation*}
Thanks to Eq. \eqref{motion}, we deduce:

\noindent -- The velocity circulation on a closed, isentropic
fluid-curve is constant.

\noindent -- In a homentropic motion (the specific entropy is
uniform in the fluid), the velocity circulation on a fluid-curve
is constant.

\noindent --  The velocity circulation on a closed fluid-curve
such that $\theta = \theta_0$ is constant.\\

\noindent From
\begin{equation*}
\boldsymbol{\gamma} -\frac{1}{2}\,\mathrm{{grad} \left ( \boldsymbol{u}^T%
\boldsymbol{u} \right) = \frac{\partial\boldsymbol{u}}{\partial t}+ \frac{%
\partial\boldsymbol{u}}{\partial \boldsymbol{x}}\  \boldsymbol{u} -\left(%
\frac{\partial\boldsymbol{u}}{\partial \boldsymbol{x}}\right)^T\boldsymbol{u}%
= \frac{\partial\boldsymbol{u}}{\partial t}+ {\rm rot}\,
\boldsymbol{u} \times \boldsymbol{u}\, .}
\end{equation*}
and
\begin{equation*}
\boldsymbol{\gamma}+{\mathrm{{grad}}\left(H+\Omega \right) -\theta
\ {\rm grad}\, s=0 \,,}
\end{equation*}
in the case of stationary motions, we obtain
\begin{equation}
{\mathrm{rot}}\, \boldsymbol{u} \times \boldsymbol{u} = \theta\,
{\mathrm{grad}}\, s - {\mathrm{grad}}\left( \frac{1}{2}\,
 \boldsymbol{u}^T\boldsymbol{u} + H+\Omega\right). \label{Crocco}
\end{equation}
Equation \eqref{Crocco} is the \emph{generalized Crocco-Vazsonyi}
relation for multi-gradient fluids.\newline The Noether theorem
proves that any law of conservation can be represented by an
invariance group. It has been proved that the conservation laws
expressed by Kelvin's theorems correspond
to group of permutations consisting of particles of equal specific entropy \cite%
{Gouin9}. It is clear that this group keeps the equation of
motions invariant both for the classical perfect fluids and also
for multi-gradient perfect fluids. Consequently, it is natural to
surmise that  the most general perfect fluids  are  continuous
media whose motions verify Kelvin's theorems.
\section{Equation of energy and thermodynamical relations}

\subsection{Equation of balance of energy}

Let us consider a dissipative fluid. The equation of motion
written in the conservative case can be extended to viscous fluids
in the  form
\begin{equation*}
\rho \, {\boldsymbol{\gamma}}+\rho \,  {\rm grad}\, \Xi +\eta \,
{\rm grad} \, \theta -{\rm div}\, {\boldsymbol{\sigma}}_{v} + \rho
\, {\rm grad}\, \Omega =0 ,
\end{equation*}
where $\boldsymbol{\sigma }_{v}$ denotes the viscous stress tensor
of the fluid. Due to the relaxation time of viscous stresses, the
viscosity  does not take into account of any gradient terms. We
denote $\boldsymbol{M}, B, N$ and $F$ as,
\begin{equation*}
\left\{
\begin{array}{l}
\displaystyle\quad\boldsymbol{M} = \rho \,
{\boldsymbol{\gamma}}+\rho \,  {\rm grad}\, \Xi +\eta \, {\rm grad}
\, \theta -{\rm div}\, {\boldsymbol{\sigma}}_{v} + \rho \,
{\rm grad}\, \Omega, \\
\\
\displaystyle\quad B = \frac{\partial \rho }{\partial
t}+\mathrm{{div}\left( \rho
\,\boldsymbol{u}\right)}, \\
\\
\displaystyle\quad N = \rho \,\theta \frac{ds}{dt}+ {\rm
div}\,\boldsymbol{Q}
- r - {\rm tr}\left(\boldsymbol{\sigma }_{v}\, D \right), \\
\\
 \displaystyle\quad F = \frac{\partial }{\partial t}\left( \frac{1}{2}\,\rho
\boldsymbol{u}^{T }\boldsymbol{u}+\rho \, \Xi +\eta \, \theta -\Pi
+\rho \,
\Omega \right) +\\
\displaystyle  {\rm div}\left\{ \left[\left( \frac{1}{2}\,\rho\,
\boldsymbol{u}^{T }\boldsymbol{u}+\rho\,{\Xi} +\eta\,\theta +\rho
\,\Omega
\right)  {\boldsymbol I} + \boldsymbol{\sigma }_{v}\right] \boldsymbol{u}+%
\boldsymbol{\chi }\right\} +{\rm
div}\,\boldsymbol{Q}-r-\rho\,\frac{\partial
\Omega }{\partial t} . %
\end{array}%
\right.
\end{equation*}
\newline Terms $\boldsymbol{Q}$ and $r$
represent the heat flux vector and the heat
supply, respectively. 
The Legendre transformation of volume internal energy
$\varepsilon$ with respect to variables $\rho, \eta, \nabla \rho,
\nabla \eta,\dots, \nabla ^{n}\rho, \nabla ^{n}\eta$ is denoted
$\Pi$,
\begin{equation}
\Pi =\rho \, \mu  +\eta \, T  +\left({\boldsymbol{\Phi }}_{1}\,
\vdots\, \nabla \rho\right) + \left({\boldsymbol{\Psi }}_{1}\,
\vdots\, \nabla \eta\right) + \ldots +\left({\boldsymbol{\Phi}}
_{n}\, \vdots\, \nabla ^{n}\rho\right) +\left({\boldsymbol{\Psi
}}_{n}\, \vdots\, \nabla ^{n}\eta\right) \, - \, \varepsilon  .
\label{pressure}
\end{equation}%
Then, $\Pi$ is a function of $\mu,  T,{\boldsymbol{%
\Phi }}_{1}, {\boldsymbol{\Psi }}_{1}, \dots,{\boldsymbol{\Phi }}_{n}, {%
\boldsymbol{\Psi }}_{n} $. \newline  Term $\displaystyle D=\frac{1}{2}\left( \partial \boldsymbol{u}%
/\partial \boldsymbol{x}+\left( \partial \boldsymbol{u}/\partial \boldsymbol{%
x}\right) ^{T }\right) $ is the velocity deformation tensor and
\begin{eqnarray*}
\boldsymbol{\chi} &=&\rho \,\frac{\partial {\boldsymbol{\Phi} _{1}}}{\partial t}+\eta \,%
\frac{\partial {\boldsymbol{\Psi} _{1}}}{\partial t} \\
&+&\left(\nabla\rho\, \vdots\,  \frac{\partial {\boldsymbol{\Phi}
_{2}}}{\partial t}\right) -\rho\,\mathrm{div} \frac{\partial
{\boldsymbol{\Phi} _{2}}}{\partial t}+\left(\nabla\eta\, \vdots\,
\frac{\partial {\boldsymbol{\Psi} _{2}}}{\partial t}\right)
-\eta\,\mathrm{div} \frac{\partial {\boldsymbol{\Psi}
_{2}}}{\partial t}+\ldots
\\
\vdots
\\
 &+&\left(\nabla ^{n-1}\rho \, \vdots\,
\frac{\partial \boldsymbol{\Phi} _{n}}{\partial
t}\right)-\left(\nabla
^{n-2}\rho \, \vdots\, \mathrm{div}\frac{\partial \boldsymbol{\Phi} _{n}}{\partial t}\right)+ \ldots    \\
&+& (-1)^{p-1}\left(\nabla ^{n-p}\rho \, \vdots\,
\mathrm{div}_{p-1}\frac{\partial
\boldsymbol{\Phi} _{n}}{\partial t}\right)+{\ldots }+(-1)^{n-1}\,\rho \,\mathrm{div}_{n-1}\frac{%
\partial \boldsymbol{\Phi} _{n}}{\partial t}   \\
&+& \left(\nabla ^{n-1}\eta \, \vdots\, \frac{\partial \boldsymbol{\Psi} _{n}}{\partial t%
}\right)-\left(\nabla ^{n-2}\eta \, \vdots\, \mathrm{div}\frac{\partial \boldsymbol{\Psi} _{n}}{%
\partial t}\right)+ \ldots    \\
&+& (-1)^{p-1}\left(\nabla ^{n-p}\eta \, \vdots\,
\mathrm{div}_{p-1}\frac{\partial
\boldsymbol{\Psi} _{n}}{\partial t}\right)+{\ldots }+(-1)^{n-1}\,\eta \ \mathrm{div}_{n-1}\frac{%
\partial \boldsymbol{\Psi} _{n}}{\partial t} .
\end{eqnarray*}
Let us note that partial derivative $\partial/\partial t$ and
${\rm div}_p$ commute. Vector $\boldsymbol{\chi}$ is similar to the
flux of energy vector obtained in \cite{Casal4}. \newline

 \emph{\textbf{Theorem}}: \emph{relation}%
\begin{equation*}
F-\boldsymbol{M}^{T }\boldsymbol{u}-\left( \frac{1}{2}\,\boldsymbol{u}^{T }%
\boldsymbol{u}+\ \Xi +s\ \theta +\ \Omega \right) \,B-N\equiv 0
\end{equation*}%
\emph{is an algebraic identity.}

\noindent Equation $\boldsymbol{M} = 0$ represents the equation of
motion, $B=0$ is the mass conservation and $N=0$ is the classical
entropy relation, then for dissipative multi-gradient fluids,
$F=0$ is the equation of energy,
\begin{equation*}
\begin{array}{c}
\displaystyle\frac{\partial }{\partial t}\left( \frac{1}{2}\,\rho
\boldsymbol{u}^{T }\boldsymbol{u} +\rho \, \Xi +\eta \, \theta
-\Pi+\rho \,
\Omega \right) + \\
\displaystyle {\rm div}\left\{ \left[ \left( \frac{1}{2}\,\rho\, \boldsymbol{u%
}^{T }\boldsymbol{u}+\rho \,\Xi +\eta \,\theta +\rho \,\Omega \right)
  {\boldsymbol  I}+{\boldsymbol{\sigma }}_{v}\right] \boldsymbol{u}+\boldsymbol{\chi }%
\right\} +{\rm div}\, \boldsymbol{Q}-r-\rho\,\frac{\partial \Omega
}{\partial t}=0 .
\end{array}%
\end{equation*}

\begin{proof}: Firstly, the proof  comes from the
identities in $\Omega $ terms,\ $(\rm{div}\, \boldsymbol{Q}-r)$
terms and $\boldsymbol{\sigma }_{v}$ terms. Secondly,  we denote
\begin{equation*}
\left\{
\begin{array}{l}
 \displaystyle\boldsymbol{M}_{0}=\rho \ \boldsymbol{\gamma}+ \rho \,\rm{grad}\,{%
\Xi }+\eta \,\rm{grad}\,\theta ,  \\
\\
 \displaystyle B=\frac{\partial \rho }{\partial t}+\rm{div}\left( \rho \,%
\boldsymbol{u}\right),  \\
\\
 \displaystyle N_{0}=\rho \,\theta \frac{ds}{dt}, \\
\\
 \displaystyle F_{0}=\frac{\partial }{\partial t}\left(
\frac{1}{2}\,\rho \boldsymbol{u}^{T }\boldsymbol{u}+\rho \, \Xi
+\eta \, \theta -\Pi \right)
+\rm{div}\left\{ \left( \frac{1}{2}\,\rho\, \boldsymbol{u}^{T }\boldsymbol{%
u}+\rho \,\Xi +\eta \,\theta \right) \boldsymbol{u}+\boldsymbol{\chi }%
\right\},
\end{array}%
\right.
\end{equation*}
then,
\begin{eqnarray*}
\rho \,\boldsymbol{\gamma}^{T }\,\boldsymbol{u} &+&\rho\,
\,\frac{\partial {\Xi
}}{\partial \boldsymbol{x}}\boldsymbol{u}+\eta \,\frac{\partial {\theta }}{%
\partial \boldsymbol{x}}\ \boldsymbol{u}=\rho \frac{d}{dt}\left( \frac{1}{2}%
\,\boldsymbol{u}^{T }\boldsymbol{u}+\Xi \right)  +\eta \frac{d\theta }{dt}%
-\rho \frac{\partial \Xi }{\partial t}-\eta \frac{\partial \theta
}{\partial
t} \\
&=&\frac{\partial }{\partial t}\left( \frac{1}{2}\,\rho \,\boldsymbol{u}%
^{T }\boldsymbol{u}+\rho \,\Xi +\eta \,\theta \right)
+\rm{div}\left[ \left( \frac{1}{2}\,\rho \,\boldsymbol{u}^{T
}\boldsymbol{u}+\rho \,\Xi
+\eta \,\theta \right) \boldsymbol{u}\right] -\rho \ \theta \frac{ds}{dt} \\
&&-\rho \frac{\partial \Xi }{\partial t}-\eta \frac{\partial \theta }{%
\partial t}-\left( \frac{1}{2}\,\boldsymbol{u}^{T }\boldsymbol{u}+\Xi
+s\ \theta \right) \left( \frac{\partial \rho }{\partial
t}+\rm{div}\left( \rho \,\boldsymbol{u}\right) \right) .
\end{eqnarray*}
Moreover,  we have
\begin{eqnarray}
\left(\nabla \rho \, \vdots\, \frac{\partial {\boldsymbol{\Phi }}_{1}}{\partial t}\right) &=&{%
\rm{div}}\left( \rho \ \frac{\partial {\boldsymbol{\Phi }}_{1}}{\partial t}%
\right) -\rho \ \frac{\partial \,{\rm{div}\boldsymbol{\Phi }}_{1}}{%
\partial t}  \notag  \\
\left(\nabla ^{2}\rho \, \vdots\, \frac{\partial {\boldsymbol{\Phi }}_{2}}{\partial t}\right) &=&%
{\rm{div}}\left[ \left(\nabla \rho \, \vdots\,   \frac{\partial {\boldsymbol{\Phi }}_{2}%
}{\partial t}\right)-\rho \ \frac{\partial\, {\rm{div}\, \boldsymbol{\Phi }}_{2}}{%
\partial t}\right] +\rho \ \frac{\partial \,{\rm{div}}_{2}{\,\boldsymbol{%
\Phi }}_{2}}{\partial t}  \notag \\
&\ \vdots\ &  \notag  \\
\left(\nabla ^{n}\rho \, \vdots\, \frac{\partial\, {\boldsymbol{\Phi }}_{n}}{\partial t}\right) &=&%
{\rm{div}}\left[ \left(\nabla ^{n-1}\rho \, \vdots\,
\frac{\partial {\boldsymbol{\Phi
}}_{n}}{\partial t}\right)-\left(\nabla ^{n-2}\rho \, \vdots\,   \frac{\partial\, {%
{\rm div} \,\boldsymbol{\Phi }}_{n}}{\partial t}\right)+{\ldots }\right. \   \notag \\
 &+&\left. (-1)^{p-1} \left(\nabla ^{n-p}\rho \, \vdots  \, \frac{%
\partial\, {\rm{div}}_{p-1} {\,\boldsymbol{\Phi }}_{n}}{\partial t}\right)+{\ldots +}(-1)^{n-1}\rho
 \,\,\frac{\partial\, {
\rm{div}}_{n-1} {\,\boldsymbol{\Phi }}_{n}}{\partial t}\right]
\notag \\
&+&(-1)^{n}\rho \ \frac{\partial \,{\rm{div}}_{n}{\,\boldsymbol{\Phi }}_{n}%
}{\partial t},  \notag
\end{eqnarray}
and a similar relation for $\eta$,
\begin{eqnarray*}
\left(\nabla \eta \, \vdots\, \frac{\partial {\boldsymbol{\Psi
}}_{1}}{\partial
t} \right)&=&\mathrm{div}\left( \eta \ \frac{\partial {\boldsymbol{%
\Psi }}_{1}}{\partial t}\right) -\eta \ \frac{\partial \,%
\mathrm{{div}\boldsymbol{\Psi }}_{1}}{\partial t}  \notag \\
\left(\nabla ^{2}\eta \, \vdots\, \frac{\partial {\boldsymbol{\Psi
}}_{2}}{\partial {t}}\right) &=&\mathrm{div}\left[ \left(\nabla \eta \,
\vdots\,   \frac{\partial {\boldsymbol{\Psi }}_{2}}{\partial
t}\right)-\eta \ \frac{\partial\, \mathrm{{div}\ \boldsymbol{\Psi
}}_{2}}{\partial t}\right]
+\eta \ \frac{\partial \,\mathrm{div}_{2}{\,\boldsymbol{\Psi }}_{2}}{%
\partial t}  \notag \\
&\, \vdots\, &  \label{lemme2} \\
\left(\nabla ^{n}\eta \, \vdots\, \frac{\partial {\boldsymbol{\Psi
}}_{n}}{\partial
t}\right) &=&\mathrm{div}\left[ \left(\nabla ^{n-1}\eta \, \vdots\,   \frac{%
\partial {\boldsymbol{\Psi }}_{n}}{\partial t}\right)-\left(\nabla
^{n-2}\eta \, \vdots\,     \frac{\partial\,{\rm div}\,
{\boldsymbol{\Psi}}_{n}}{\partial t}\right)+ \ldots   \right.     \notag \\
&+&\left. (-1)^{p-1}\left(\nabla ^{n-p}\eta \, \vdots\,   \frac{%
\partial\,  {\rm div}_{p-1}\, {\boldsymbol{\Psi }}_{n}}{\partial
t}\right)+ \ldots +
(-1)^{n-1}\eta\, \,\frac{{\partial\,\rm div}_{n-1}\, {\boldsymbol{\Psi }}_{n}%
}{\partial t}\right]  \notag \\
&+&(-1)^{n}\eta \ \frac{\partial \,\mathrm{div}_{n}{\,\boldsymbol{\Psi }}_{n}%
}{\partial t}.  \notag
\end{eqnarray*}
From  Eqs. \eqref{differentielenergy} and \eqref{pressure}, we
deduce
\begin{equation*}
d\Pi = \rho \, d\mu   +\eta \, dT  +\left(\nabla \rho \, \vdots\,
d{\boldsymbol{\Phi }}_{1}\right) +  \left(\nabla \eta\, \vdots\,
d\boldsymbol{\Psi }_{1}\right)  + \ldots +\left(\nabla
^{n}\rho\, \vdots\, d\boldsymbol{\Phi}%
_{n}\right)  +\left(\nabla ^{n}\eta \, \vdots\, d{\boldsymbol{\Psi
}}_{n}\right)
\end{equation*}
which implies
\begin{equation*}
 \frac{\partial\Pi}{\partial t} =\rho \, \frac{\partial \mu}{\partial t}  +
 \eta \, \frac{\partial T}{\partial t} +\left(\nabla \rho \,
 \vdots\,
 \frac{\partial{\boldsymbol{\Phi }}_{1}}{\partial t}\right)+
\left(\nabla \eta\, \vdots\,  \frac{\partial{\boldsymbol{\Psi
}}_{1}}{\partial t}\right) + \ldots +\left(\nabla ^{n}\rho\,
\vdots\, \frac{\partial{\boldsymbol{\Phi }}_{n}}{\partial
t}\right)+\left(\nabla ^{n}\eta\, \vdots\,
\frac{\partial{\boldsymbol{\Psi }}_{n}}{\partial t}\right).
\end{equation*}
Partial derivative $\partial/\partial t$ and ${\rm div}_p$ commute
and consequently,
\begin{eqnarray*}
\displaystyle\frac{\partial \Pi }{\partial t} &=&\rho \frac{\partial \mu }{%
\partial t} -\rho \ \frac{\partial \,{%
\rm{div}\boldsymbol{\Phi }}_{1}}{\partial \boldsymbol{x}} +{\ldots }%
+(-1)^{n}\rho \ \frac{\partial \,{\rm{div}}_{n}{\,\boldsymbol{\Phi }}_{n}}{%
\partial t}   \\
\displaystyle  &+&   \eta \frac{\partial T}{\partial t}  -\eta \ \frac{%
\partial \,{\rm{div}\boldsymbol{\Psi }}_{1}}{\partial t}+{\ldots }%
 +(-1)^{n}\eta \ \frac{\partial \,{\rm{div}}_{n}{\,\boldsymbol{%
\Psi }}_{n}}{\partial t}
\\
&+&\displaystyle   {\rm{div}}\ \left[\, \rho \ \frac{\partial
{\boldsymbol{\Phi
}}_{1}}{\partial t}+ \ldots  +\left(\nabla ^{n-1}\rho \, \vdots\,   \frac{\partial {%
\boldsymbol{\Phi }}_{n}}{\partial t}\right)-\left(\nabla ^{n-2}\rho \, \vdots\, \rm{div}%
\frac{\partial {\boldsymbol{\Phi }}_{n}}{\partial t}\right)+{\ldots } \right.  \\
&+&\displaystyle(-1)^{p-1}\left(\nabla ^{n-p}\rho \, \vdots\,   {\rm{div}}_{p-1}\frac{%
\partial {\boldsymbol{\Phi }}_{n}}{\partial t}\right)+ \ldots + (-1)^{n-1}\rho \, {%
\rm{div}}_{n-1}\frac{\partial {\boldsymbol{\Phi }}_{n}}{\partial t} \\
&+&\displaystyle\eta \ \frac{\partial {\boldsymbol{\Psi }}_{1}}{\partial t}+{%
\ldots }+\left(\nabla ^{n-1}\eta \, \vdots\,   \frac{\partial {\boldsymbol{\Psi }}_{n}}{%
\partial t}\right)-\left(\nabla ^{n-2}\eta \, \vdots\,   \rm{div}\frac{\partial {\boldsymbol{%
\Psi }}_{n}}{\partial t}\right)+ \ldots   \\
&+&\displaystyle\left. (-1)^{p-1}\left(\nabla ^{n-p}\eta \, \vdots\,   {\rm{div}}_{p-1}%
\frac{\partial {\boldsymbol{\Psi }}_{n}}{\partial t}\right)+
\ldots +
(-1)^{n-1}\eta \,\,{\rm{div}}_{n-1}\frac{\partial {\boldsymbol{\Psi }}_{n}%
}{\partial t}\ \right],
\end{eqnarray*}%
and from \eqref{tempchem}, we get
\begin{equation*}
\frac{\partial \Pi }{\partial t}=\rho \frac{\partial \Xi
}{\partial t}+\eta \frac{\partial \theta }{\partial
t}+\rm{div}\,\boldsymbol{\chi }.
\end{equation*}%
Then,
\begin{eqnarray*}
\displaystyle\boldsymbol{M}_{0}^{T }\,\boldsymbol{u} &\boldsymbol{=}&%
\frac{\partial }{\partial t}\left( \frac{1}{2}\,\rho
\,\boldsymbol{u}^{T }\boldsymbol{u}+\rho \,\Xi +\eta \,\theta -\Pi
\right) +\rm{div}\left[ \left( \frac{1}{2}\,\rho
\,\boldsymbol{u}^{T }\boldsymbol{u}+\rho \,\Xi
+\eta \,\theta \right) \boldsymbol{u+\chi }\right]  \\
&&\displaystyle-\rho \ \theta \frac{ds}{dt}-\left( \frac{1}{2}\,\boldsymbol{u%
}^{T }\boldsymbol{u}+\Xi +s\ \theta \right) \left( \frac{\partial \rho }{%
\partial t}+\rm{div}\left( \rho \,\boldsymbol{u}\right) \right),
\end{eqnarray*}%
which implies%
\begin{equation*}
F_{0}-\boldsymbol{M}_{0}^{T }\boldsymbol{u-}\left( \frac{1}{2}\,%
\boldsymbol{u}^{T }\boldsymbol{u}+\Xi +s\,\theta   \right)
B-N_{0}\equiv 0
\end{equation*}%
and proves the theorem. \end{proof}

\subsection{The Planck and the Clausius-Duhem  inequalities}

For all dissipative fluid motions,
\begin{equation*}
{\rm tr}\left( \boldsymbol{\sigma }_{v}D\right)\geq 0.
\end{equation*}
Then, from relation $N = 0$, we deduce the Planck inequality as in
\cite{Truesdel1},
\begin{equation*}
\rho \,\theta\, \frac{ds}{dt}+\mathrm{{div}\,\boldsymbol{Q} - r
\geq 0.}
\end{equation*}
We consider  the Fourier inequality in the   general form
\begin{equation*}
\boldsymbol{Q}^{T}\mathrm{grad}\, \theta \leq 0
\end{equation*}
and we obtain
\begin{equation*}
\rho \,\frac{ds}{dt}+\mathrm{{div}\,\frac{\boldsymbol{Q}}{\theta }-\frac{r}{%
\theta } \geq 0,}
\end{equation*}
which is the extended form of the Clausius-Duhem inequality.
\newline
Consequently, multi-gradient fluids are compatible with the first
and second laws of thermodynamics.
\newline

\noindent {\emph{Final remark}}: In a forthcoming article, we will
prove that system of equations of multi-gradient fluids is a
quasi-linear hyperbolic-parabolic system of conservation laws
which can be written in  Hermitian symmetric-form implying the
stability of constant solutions.
\newline

 \noindent {\footnotesize{\textbf{Acknowledgments:}
{The results contained in the present paper
have been partially presented in Wascom 2017.}} }

\end{document}